\documentclass[twoside,10pt]{article}
\usepackage{amsmath,amssymb}
\usepackage{xspace}
\usepackage{tam13}

\usepackage{undertilde,citesort}
\usepackage{amscd} 

\title{ Time and Fermions:  General Covariance \emph{vs.} Ockham's Razor for Spinors  }
\shorttitle{ Time and Fermions}
\authors{J.~B.~Pitts$^1$\email{jbp25@cam.ac.uk}}
\shortauthors{J.~B.~Pitts}
\affiliations{
$^1$Faculty of Philosophy, Sidgwick Avenue, University of Cambridge, Cambridge, United Kingdom CB3 9DA
}

\abstract{
It is a commonplace in the foundations of physics, attributed to Kretschmann, that any local physical theory can be represented using arbitrary coordinates, simply by using tensor calculus.  On the other hand, the physics and mathematics literature often claims  that spinors \emph{as such} cannot be represented in coordinates in a curved space-time.  These commonplaces are inconsistent.  What general covariance means for theories with fermions is thus unclear.

In fact both commonplaces are wrong.  Though it is not widely known, Ogievetsky and Polubarinov (OP) constructed spinors in coordinates in 1965, enhancing the unity of physics and helping to spawn particle physicists' concept of nonlinear group representations.  Roughly and locally, OP spinors resemble the orthonormal basis or tetrad formalism in the symmetric gauge, but they are conceptually self-sufficient and more economical.  The typical tetrad formalism is thus de-Ockhamized, with six extra field components and six compensating gauge symmetries to cancel them out.  As developed nonperturbatively by Bilyalov, OP spinors admit any coordinates at a point, but `time' must be listed first; `time' is defined in terms of an eigenvalue problem involving the metric components and the matrix $diag(-1,1,1,1)$, the product of which must have no negative eigenvalues.  Thus even formal general covariance requires reconsideration; the atlas of admissible coordinate charts should be sensitive to the types and \emph{values} of the fields involved.

Apart from coordinate order and the usual spinorial two-valuedness, (densitized) Ogievetsky-Polubarinov spinors form, with the (conformal part of the) metric, a nonlinear geometric object.  Important results on Lie and covariant differentiation are recalled and applied.  The rather mild consequences of the coordinate order restriction are explored in two examples: the question of the conventionality of simultaneity in  Special Relativity, and the Schwarzschild solution in General Relativity.  
}

\begin{document}
\maketitle

\section{Introduction  }

During 1929-37 especially due to Weyl and Cartan,  it was concluded that spinor fields in curved space-time could not be treated using the techniques of tensor calculus;  a spinorial transformation law had to be achieved with respect to some group unrelated to spatial rotations, with the help of an orthonormal basis  \cite{WeylGravitationElectron,WeylElektronGravitation,WeylRice,WeylGroups,CartanSpinor}.
The no-go  theorem in question is usually taken to show that one cannot have spinors in a curved space-time  with spinorial behavior under coordinate transformations. This result was considered foundational and permanent, not technical and provisional   \cite{ScholzWeylFockSpinor}.

During the 1950s-60s it was found how to circumvent the Weyl-Cartan result:  spinors could be spinorial with respect to coordinate transformations if one allowed the new spinor components also depend nonlinearly on the metric.  But this result has yet to become widely known.  The DeWitts provided clues  in the early 1950s \cite{DeWittSpinor}. Pauli might also have had a role \cite{BelinfanteDiracSpinorCommutation}.
The problem was solved, at least infinitesimally, by V. I. Ogievetsky and I. V. Polubarinov \cite{OPspinorReprint,OP}. 
This work, which largely began particle physicists' idea of nonlinear group representations, unfortunately never made contact with the mathematics literature on nonlinear geometric objects, such as (\cite{Tashiro1,Nijenhuis,Tashiro2,SzybiakCovariant,SzybiakLie}).  While it isn't possible to write this paper in the 1960s, one can imagine a different trajectory for certain literatures in the last 40+ years if that connection had been made.   As Ogievetsky and Polubarinov (henceforth ``OP'') said in critique of Cartan, the Weyl-Cartan no-go result relies essentially on largely tacit assumptions, such as that the new spinor components should be linear in the old field components and that the old field components should include only the spinor itself.\footnote{The content of this paper is explored in more detail in \cite{PittsSpinor}.  }

One can approach  OP  formalism by gauge-fixing the orthonormal basis to make its components symmetric, moving an index with the signature matrix $diag(-1,1,1,1)$.    For an indefinite signature, this gauge fixing doesn't always work.  But it works often enough that one might decide to live it, given the benefits of avoiding 6 extra fields and 6 extra symmetries (local $O(1,3)$) to gauge them away, and having Lie and covariant derivatives with resulting benefits for symmetries and conservation laws.

Strikingly, a great many authors have in effect reinvented the OP formalism by such gauge fixing---generally without realizing that a formalism that could stand on its own was thereby achieved, and often while even denying that such a self-sufficient formalism was possible. 
  The fact that an orthonormal basis can be topologically obstructed on manifolds with a metric shows that the symmetric square root of the metric is conceptually independent of a tetrad.  As simple a case as the 2-sphere with positive definite metric makes the point.  

\subsection{Generalized Polar Decomposition}

One can obtain the symmetric square root of the metric using a generalized polar decomposition of the tetrad, treated as a matrix of components.   The usual polar decomposition factors a positive definite matrix into an orthogonal (rotation) factor and a symmetric (shear and expansion) factors.  (Clearly the expansion, which corresponds to the volume element in our applications, goes along for the ride and can be separated out using a unimodular matrix and a scalar density factor.) 
Isham, Salam and Strathdee have invoked the polar decomposition near the identity, but without mathematical control over what happens for large transformations \cite{IshamSalamStrathdee,IshamSalamStrathdee2}.  
A recent paper in the modern style is also noteworthy \cite{GiachettaSpinor}, though still working near the identity.  There might be some interesting connection to Ol'shanki\u{i} decompositions \cite{LawsonPolarOlshanskii}.  In any case progress in  linear algebra has  filled the hole.  

A generalized polar decomposition makes use of a ``signature matrix'' like $\eta=diag(-1,1,1,1)$ \cite{Bolshakov1,Higham}. A matrix $M$ is $\eta$-orthogonal iff $M^T \eta M = \eta$.  It is called $\eta$-symmetric if symmetric with index moved by $\eta$. 
 Tweaking some  notation a bit, one has the following theorem applicable to generalized polar decomposition of a tetrad into a symmetric square root and a boost-rotation: 
\begin{quote} Theorem 5.1. If [tetrad component matrix] $E \in  R^{n \times n}$ and $\eta E^{T}\eta E$ has no eigenvalues on the nonpositive real axis, then $E$ has a unique indefinite polar decomposition $E = QS$, where $Q$ is
$\eta$-orthogonal and $S$ is $\eta$-symmetric with eigenvalues in the open right half-plane. \cite[p. 513]{Higham} 
\end{quote}
One can find the boost-rotation explicitly:   $Q=E(\eta E^T \eta E)^{-\frac{1}{2}}$.
Having made this decomposition of the tetrad $E,$ the OP formalism works with $S$ or $R = \eta S,$ throwing away the local $O(1,3)$ factor, which is pure gauge, and keeping the symmetric factor, the symmetric square root of the metric, a physical quantity equivalent to the metric but more directly coupled to the spinor in the Dirac equation.

\subsection{General Covariance and Coordinate Reordering}

General coordinate transformations contain a certain collection of transformations that probably  no one has ever regarded as interesting prior to Ranat Bilyalov's work on spinors \cite{BilyalovConservation}.  
The usual formalisms require that the coordinates be expressed in some definite order, but no significance is attached to that order. Consider  Special Relativity in Cartesian coordinates. Coordinates are listed as an \emph{ordered} quadruple  $(t, x, y, z)$ with their usual meanings.  A slightly nontrivial change is a reordering that involves the time coordinate:  $(t, x, y, z)$ might be replaced by $(x, t, y, z).$ Now the  matrix $diag(-1,1,1,1)$ is inappropriate; the invariant interval in the new coordinates is given by $diag(1,-1,1,1).$   In short, one needs tensor calculus, or a small fragment  of it, namely the part that introduces a metric tensor and its coordinate transformation law, in order to permit the coordinate reordering. One certainly could formulate Special Relativity in this fashion.  Presumably no one would bother to introduce a metric tensor simply for the slight gain in generality of admitting reorderings of the Cartesian coordinates, however.  The indefinite signature of the metric, which distinguishes temporal from spatial coordinates, is crucial here.
Roughly, the freedom to write `time' (suitably generalized) as the second, third or fourth coordinate (in four space-time dimensions) is  what one \emph{gives up} in the OP spinor formalism 
 \cite{BilyalovSpinors}.  
Contrary to the usual modern practice of admitting all possible coordinates, it is fitting to take the atlas of admissible coordinate charts to vary with the types of fields present in order to make a lean ontology manifest.

It is helpful to remember an argument that Einstein made in 1916 when the issue was novel: 
\begin{quote}  The method hitherto employed for laying co-ordinates into the space-time continuum in a definite manner [yielding observable time or space intervals] thus breaks down, and there seems to be no other way which would allow us to adapt systems of co-ordinates to the four-dimensional universe so that we might expect from their application a particularly {\bf simple} formulation of the laws of nature. So there is nothing for it but to regard all imaginable systems of co-ordinates, on principle, as equally suitable for the description of nature. 
\cite[p. 117, emphasis added]{EinsteinFoundation} \end{quote}
Indeed coordinates with \emph{quantitative} physical meaning are not available with adequate generality in the curved space-times of GR.  The next crucial step involves inferring the nonexistence of something, given the failure thus far (in 1916) to imagine it.  This step was eminently reasonable in 1916 and indeed for some time afterward.  But clearly the argument is defeasible, and now it is defeated.

\section{Geometric Objects, Especially Nonlinear Ones}

The  theory of geometric objects, which was largely complete in the 1960s in the linear case, 
describes tensors and more general geometric objects, including connections, as well as some little known nonlinear entities, some but not all of which are equivalent to linear ones.  
This theory, which was never well known, is even less so now (except for some work  on ``natural bundles''   \cite{NijenhuisYano,FatibeneFrancaviglia}).

Roughly, a geometric object is, for each space-time point and each local coordinate system around it, a finite ordered set of   components and a transformation law relating components in different coordinate systems at the same point \cite{Nijenhuis,KucharzewskiKuczma,Trautman}.  
Tensors have linear homogeneous transformation laws: $v^\prime \sim v,$ where the usual flurry of indices is suppressed, partly for generality.
A linear homogeneous transformation law also exists for  tensor densities \cite{Anderson}, for which the transformation law involves some power of the determinant of the matrix $\frac{ \partial x^{\mu^\prime} }{\partial x^{\nu}  } .$
An affine connection is a geometric object with a linear but inhomogeneous (affine) transformation law.   These are all the geometric objects in wide circulation, but others exist, such as the metric perturbation $g_{\mu\nu} - \eta_{\mu\nu}$ (where $\eta_{\mu\nu}=diag(-1,1,1,1)$) \cite{OP,OPspinorReprint}.  It  has an affine transformation law with only first derivatives, so it has nice Lie and covariant derivatives \cite{Yano}.  It should be emphasized that $\eta_{\mu\nu}=diag(-1,1,1,1)$) is a purely numerical scalar matrix.  Somewhat awkwardly as far as notation is concerned, the theory of nonlinear geometric objects, linear for a subgroup, involves cases where Greek coordinate indices and Latin indices merge \cite{OPspinorReprint,ChoFreund}. 



Besides linear and affine geometric objects, there are  geometric objects with nonlinear transformation laws: the new components are \emph{nonlinear} functions of the old components, as well as depending on  $\frac{ \partial x^{\mu^\prime} }{\partial x^{\nu}  }(p)$ and the like.   Consider the symmetric square root  $r_{\mu\nu}$ of the metric,\footnote{The square root of the metric or inverse metric has become important recently in devising pure spin 2 massive gravity \cite{HassanRosen}.  Some of the theories in question were first developed by OP \cite{OP}, though OP did not notice or resolve the problem \cite{DeserMass} that getting rid of the ghost spin 0 by making it infinitely heavy in linearized massive gravity might not get rid of it to all orders.}  defined implicitly in any (admissible) coordinate system by  
\begin{equation} 
g_{\mu\nu} = r_{\mu\alpha} \eta^{\alpha\beta} r_{\beta\nu}.
\end{equation}
 This entity $r_{\mu\nu}$ exists at least  in many coordinate systems, most obviously in those not terribly far from freely falling Cartesian coordinates.  If one expresses the metric $g_{\mu\nu}$ as some perturbation about the matrix $\eta_{\alpha\beta}=diag(-1,1,1,1),$ then one can use the binomial series expansion for  $r_{\mu\alpha}.$ The result (when convergent \cite{DeWittSpinor}) is \cite{OPspinorReprint}
\begin{eqnarray} r_{\mu\nu} =
 \sum_{k=0}^{\infty} \frac{ \frac{1}{2}! }{ (\frac{1}{2}-k)! k! } [(g_{\mu\bullet}-\eta_{\mu\bullet}) \eta^{\bullet\bullet}\ldots (g_{\bullet\nu}-\eta_{\bullet\nu})]^{k\ factors \ of \ g} \nonumber \\ = \eta_{\mu\nu} + \frac{1}{2}(g_{\mu\nu}-\eta_{\mu\nu}) - \frac{1}{8}(g_{\mu\alpha}-\eta_{\mu\alpha}) \eta^{\alpha\beta} (g_{\beta\nu}-\eta_{\beta\nu}) + \ldots. \label{Series} \end{eqnarray}
The expression $\frac{1}{2}!/ (\frac{1}{2}-k)! $ stands for $\frac{1}{2} \cdot (\frac{1}{2} -1) \cdot \ldots \cdot(\frac{1}{2}-k + 1).$  The coordinate transformation law for $r_{\mu\alpha}$ follows from the metric transformation law for $g_{\mu\nu}$  and the definition as applied to both coordinate systems: 
$g=r \eta r$  and $g^{\prime} = r^{\prime} \eta r^{\prime}:$  
$$r_{\mu\alpha}^{\prime} \eta^{\alpha\beta} r_{\beta\nu}^{\prime} = \frac{\partial x^{\alpha} }{ \partial x^{\mu\prime} } r_{\alpha\rho} \eta^{\rho\sigma} r_{\sigma\beta} \frac{ \partial x^{\beta} }{ \partial x^{\nu\prime} }.$$  
  This result is somewhat implicit in that $r_{\mu\alpha}^{\prime}$ does not appear alone on the left side. The matrix $diag(-1,1,1,1)$ puts time first; if one tries to order the coordinates otherwise, then either one has to replace the matrix $diag(-1,1,1,1)$ with something else  or one gets  perturbations of magnitude $\pm 2$.   Bilyalov presents a \emph{theorem} \cite{BilyalovConservation} regarding the necessity \emph{and sufficiency} of reordering the coordinates, not simply a plausibility argument for its necessity.  The coordinate reordering is in effect part of the service rendered by his  matrix $T$, which combines a permutation and a reflection.


It is evident from the definition of the symmetric square root of the metric that there is a transformation rule for it, at least between admissible coordinate systems.  That fact is illustrated in the `commutative diagram' (with factors of $diag(-1,1,1,1)$ suppressed):
$$\begin{CD}
g^{\prime} @<tensor << g\\
@VrootVV @VrootVV \\
r^{\prime} @<?<< r
\end{CD}$$
But what can be said that is explicit and practical about the transformation from $r_{\mu\nu}$ to $r^{\prime}_{\mu\nu}$, labeled as ``?'' in the diagram? 
It is   \begin{eqnarray}
r^{\prime}_{\mu\nu} =  \sqrt{  \frac{ \partial x^{\alpha} }{ \partial x^{\mu\prime} }   r_{\alpha\beta} \eta^{\beta\gamma} r_{\gamma\rho} \frac{ \partial x^{\rho} }{ \partial x^{\sigma\prime} } \eta^{\sigma\delta}  } \eta_{\delta\nu}. \end{eqnarray} 
When the perturbative expansion exists, one can  use $\left(\frac{  \partial x}{ \partial x^{\prime} } \right)^\intercal  r \eta r \frac{ \partial x }{ \partial x^{\prime} } \eta - I$ as the perturbation in the binomial series expansion. 
 Thus the transformation rule, at least in the perturbative context, is an infinite series in even powers of $r$.  The series expansion chooses the root near the identity, that is, near $diag(-1,1,1,1);$ presumably one  wants  the `positive' principal square root in all other contexts also. Bilyalov's generalized eigenvector formalism  works more  generally \cite{BilyalovSpinors}, but still requires that  `time' be listed  first among the coordinates; otherwise negative eigenvalues can appear, yielding a complex square root of the metric.  


\subsection{Conformal  Group Yields Linear Transformation Law}

Besides the full nonlinear transformation law for (nearly) general coordinate transformations, it is of interest to ascertain when  the transformation law is linear---the stabilizer group.  The answer is the 15-parameter  conformal group. (At times only the Poincar\'{e} group, not the full 15-parameter conformal group, has been noticed \cite{OPspinorReprint}.) One can show that infinitesimally using the Lie derivative of OP spinors, as given in (\cite{OPspinorReprint}).  One can also show it using the finite coordinate transformation law. No  special assumptions about the metric geometry or the coordinate system are made; the transformations involve purely formal relations between a new coordinate system and an old one, regardless of the metric or the metrical properties of the coordinates.  
Thus the simple linear spinor-only law for Lorentz transformations in quantum field theory books arises as a special case.

\subsection{Differentiation of Nonlinear Geometric Objects and OP Spinors }

For a nonlinear geometric object $\chi,$ generally the Lie and covariant derivatives themselves are not geometric objects, but the pairs  $\langle \chi, \mathcal{L}_{\xi} \chi \rangle$  and
 $\langle \chi, \nabla  \chi \rangle$ are both geometric objects \cite{Tashiro1,Tashiro2,Yano,SzybiakCovariant,SzybiakLie}. 
One often reads that spinors do not admit Lie differentiation, unless the generating vector field is a conformal Killing vector  field \cite{BennTucker} \cite[p. 101]{PenroseRindler2}.  
OP spinors do in fact have a classical Lie derivative, as one would expect from their having a spinorial coordinate transformation law and no additional gauge group. One in fact must consider a pair  such as   $\langle r_{\mu\nu}, \psi  \rangle$ as a candidate for Lie differentiation. Thus   $\langle r_{\mu\nu}, \psi,  \mathcal{L}_{\xi} r_{\mu\nu}, \mathcal{L}_{\xi} \psi \rangle$ is a geometric object (apart from coordinate restrictions and spinorial double-valuedness). 
Regarding the covariant derivative of the metric, one has  that  $\nabla g_{\mu\nu}=0  \leftrightarrow \nabla r_{\mu\nu}=0.$ Thus the covariant derivative of OP spinors simplifies somewhat due to the disappearance of $\nabla r_{\mu\nu}$ terms.

\section{Conformal \emph{In}variance of Dirac Operator with Densities}

One common theme in modern physics is the value of identifying and eliminating surplus structure  \cite{Anderson,TLL,LLN,FriedmanJones}.
There are two  distinct but compatible ways of eliminating surplus structure from the typical formulation of the massless Dirac equation in a curved space-time.  One entity that can be eliminated, as was realized by Haantjes and Schouten in the 1930s \cite{SchoutenHaantjesConformal,HaantjesConformalSpinor,PeresPolynomial} but mostly forgotten afterwards, is the volume element $\sqrt{-g}.$  Many authors nowadays discuss the \emph{co}variance of the Dirac operator (the left side of the massless Dirac equation) under conformal changes of metric, but  one can, using a suitable choice of primitive fields, achieve \emph{in}variance, in which $\sqrt{-g}$ disappears altogether from the theory.  
Failure to notice leads some authors to think that Killing vectors only  \cite{KosmannLie,KosmannLieApplied,FatibeneFermion,Cotaescu,FatibeneFrancaviglia}, not conformal Killing vectors, have a special status in relation to Lie derivatives of spinors.

It is perhaps novel to combine the elimination of the tetrad's surplus structure (6 components) with the elimination of the volume element, 1 component of surplus structure.  One benefit is that it is clear \emph{by inspection} that the symmetry group in (conformally) flat space-time is the 15-parameter conformal group.  
The appropriate spinor turns out to have weight $\frac{3}{8}$ in four space-time dimensions or, more generally, $\frac{n-1}{2n}$ in $n$ space-time dimensions. That conformally invariant Dirac operator is  $\gamma_{\mu} \hat{r}^{\mu\nu} \nabla_{\nu} \psi_{w},$ where $\gamma_{\mu}$ denotes a set of numerical Dirac matrices, $\hat{r}^{\mu\nu}$ is the symmetric square root of the inverse conformal metric density $\hat{g}^{\mu\nu},$  $\nabla_{\nu}$ is the OP covariant derivative for spinors  with the density weight term (with the weight  altered to match the usual western rather than Russian  conventions), and $\psi_{w}$ is a spinor with weight $w = \frac{n-1}{2n}.$  No use is made of any volume element in defining this operator, so it is a concomitant of just the weighted spinor and the conformal metric density. 
   The Lagrangian density in this formalism with the densitized variables is also manifestly conformally invariant in any dimension:  
\begin{equation} \mathcal{L} = \sqrt{-g} \bar{\psi} \gamma_{\nu} r^{\nu\mu} \nabla_{\mu} \psi    =   \bar{\psi}_{w} \gamma_{\nu} \hat{r}^{\nu\mu} \nabla_{\mu} \psi_{w}. 
\end{equation}


\section{Spinors and the Partial Conventionality of Simultaneity}

The conventionality of simultaneity has been a longstanding issue in the philosophy of physics. 
The question has arisen whether spinor fields pose distinctive issues for the conventionality of simultaneity.  
Detaching spinorhood from spatial coordinate rotations is curious  \cite[pp. 150, 151]{CartanSpinor} \cite{BainSpinorSimultaneity}---but an appropriate response to the supposed impossibility of including spinors as such within the realm of coordinate transformations.

 Due to OP, that impossibility is overcome.  Cartan's and Bain's preference for unification, which they thought unsatisfiable,  is adequately realized using OP spinors, for which  the group that makes  $\psi$ a spinor is the (double cover of) a Lorentz group \emph{that is a subgroup of the space-time coordinate transformations}.  
OP spinors are  friendly to conventionalism about simultaneity, but not for reasons that have appeared previously. 
   A great variety of time coordinates is permitted, including all of them usually considered in the debate over the conventionality of simultaneity debate and some that are not.  That is not, however, because any coordinate whatsoever (nor even any `flat' one linearly related to standard coordinates in Minkowski space-time) is admissible as time; in fact some coordinates are inadmissible as time coordinates in the OP spinor formalism.    The dividing line is apparently unprecedented. 
Let the transformation from standard to nonstandard simultaneity coordinates be given by $x^{\mu^{\prime} } = (x^0 + (2\epsilon_1 -1)x^1 + (2\epsilon_2 -1) x^2 + (2\epsilon_3 -1)x^3, x^1, x^2, x^3).$ Standard simultaneity is the case $\epsilon_1=\epsilon_2=\epsilon_3=\frac{1}{2}.$  It is convenient to define $\vec{n}=(2\epsilon_1 -1, 2\epsilon_2 -1, 2\epsilon_3 -1),$ so then
$x^{\mu^{\prime} } = x^{\mu} + \delta^{\mu}_0 n_i x^i.$ One finds the inverse metric for Minkowski space-time  with nonstandard simultaneity to be
\begin{eqnarray}
\left[
\begin{array}{cccc}
-1+\vec{n}^2  & n_1 & n_2 & n_3 \\ 
n_1  & 1 & 0 & 0 \\
n_2 & 0 & 1 & 0 \\
n_3 & 0 & 0 & 1 
\end{array}
\right]_.
\end{eqnarray}
 It is convenient to rotate coordinates to let $\vec{n}$ lie along one coordinate axis, and then suppress the two trivial dimensions.  Using an eigenvalue formalism, one sees that for $|n| \geq 2$, the eigenvalues are real and \emph{negative}, which is bad.  Coordinates  with $|n| < 2$ are permitted. 
 The final result is
\begin{eqnarray}
r^{\prime\mu\nu}= 
\left[
\begin{array}{cc}
-\left(1-\frac{n^2}{2}\right) \left(1-\frac{n^2}{4}\right)^{-\frac{1}{2}}   & \frac{n}{2} \left(1-\frac{n^2}{4}\right)^{-\frac{1}{2}}  \\ 
  \frac{n}{2} \left(1-\frac{n^2}{4}\right)^{-\frac{1}{2}}&  \left(1-\frac{n^2}{4}\right)^{-\frac{1}{2}}
\end{array}
\right]  \nonumber \\%
 = %
\left[
\begin{array}{cc}
-1 + \frac{3n^2}{8}+\ldots    & \frac{n}{2} + \ldots  \\ 
  \frac{n}{2} + \ldots & 1+\frac{n^2}{8} + \ldots
\end{array}
\right]_.  
\end{eqnarray} 
Recalling that $n=2\epsilon -1$  and that the usual range of $\epsilon$ in discussions of the conventionality of simultaneity is between $0$ and $1$,  the usual range of $n$ is between $-1$ and $1.$  The OP formalism permits $-2 < n < 2,$ considerably larger than the typical $-1<n<1$ of the conventionality of simultaneity discussion, but less than the full  Kretschmannian arbitrariness of  tensor calculus.  
The corresponding transformation of the spinor is 
\begin{eqnarray}
S= I \cosh \left[\frac{1}{2} \ln \sqrt{\frac{1-n/2}{1+n/2} }\right] + \gamma_0 \gamma^1 \sinh \left[\frac{1}{2} \ln \sqrt{\frac{1-n/2}{1+n/2} }\right]  \nonumber \\  %
= %
\frac{1}{2}I \left[   \sqrt[4]{\frac{1-n/2}{1+n/2} }  +    \sqrt[4]{\frac{1+n/2}{1-n/2} } \right]       + \frac{1}{2}\gamma_0 \gamma^1   \left[   \sqrt[4]{\frac{1-n/2}{1+n/2} }  -    \sqrt[4]{\frac{1+n/2}{1-n/2} } \right] \nonumber \\ %
= 
I - \frac{n}{4} \gamma_0 \gamma^1 + \ldots.    
\end{eqnarray}
A coordinate transformation from standard simultaneity to nonstandard simultaneity induces an $n$-dependent  boost of the spinor.


\section{The Schwarzschild Radius and  `Time' Coordinates}

Part of the lore of general relativists is the role of Eddington-Finkelstein coordinates in the late 1950s in helping  to overcome `Schwarzschild singularity' at $r=2M;$ instead that radius came to be known as the ``horizon'' of a black hole, through which one might readily enough pass on the way to the  curvature singularity at $r=0$ \cite{MTW}. One might well conclude that holding too tightly to an association between a coordinate and some qualitative temporal or spatial character played a negative role in the context of discovery for the significance of $r=2M,$ a role only overcome with difficulty using infalling coordinates that made no such qualitative associations. Does the  OP  formalism  un-learn lesson of Schwarzschild radius?

The infalling Eddington-Finkelstein coordinates are a radial coordinate $r$, a null coordinate $\tilde{V}$, and two angles $\theta,$ $\phi.$ The line element is given by
$$ds^2 = -\left(1- \frac{2M}{r} \right) d\tilde{V}^2  + 2  d\tilde{V}dr + r^2(d\theta^2 + sin^2\theta d\phi^2)$$ \cite[p. 828]{MTW}. 
A null coordinate such as $\tilde{V}$ is, in some rough sense, half spatial and half temporal. $r$ seems quite unambiguously spatial.  One can use the generalized eigenvalue formalism to ascertain the coordinate ranges of admissibility for these coordinates.  The results are surprising in more than one respect.  One can show that the  OP admissibility range of $\langle \tilde{V}, r, \theta, \phi \rangle$ (noting the \emph{order}) is  $r> \frac{2M}{3}$. That seems plausible enough---something strange happens somewhere inside the Schwarzschild radius, but at least one can get inside it before having to switch coordinates.  That fact alone indicates that the OP `time' coordinate restrictions do not unlearn the lessons about coordinates from 1958 that contributed to the modern understanding of black holes.  At any rate one can get inside the horizon, and there is no reason to assume that  $r> \frac{2M}{3}$ is a real  barrier.

One can also consider the  OP admissibility of $\langle r, \tilde{V}, \theta, \phi \rangle,$ with $r$ coming first. This is, intuitively, the `wrong' order, because a spatial coordinate is playing the role of OP `time,' while a null (half time, half space) coordinate is playing the role of space.  One might expect this coordinate system not to be admissible, or to be admissible only in some small exotic region, such as inside the horizon.  But on doing the calculation, one finds these coordinates to be admissible for  $r>0$!  The eigenvalues are complex.  As noted above, complex eigenvalues cause no trouble because they permit a real square root with eigenvalues in the right half of the complex plane.  Thus for  $r> \frac{2M}{3}$ there is more than one right order:  $\langle \tilde{V}, r, \theta, \phi \rangle$  and  $\langle r, \tilde{V},  \theta, \phi \rangle$ are both admissible.  The relation between the spinor components in the two systems is presumably quite nontrivial, in contrast to the relationship between the metric tensor components.  

In short, there are cases in which more than one right order exists, and cases where an intuitively wrong order is admissible and an intuitively more right order is inadmissible.   The lesson of the  Schwarzschild radius is not unlearned by OP spinors.


\section{Conclusion}

It is usually tacitly assumed that most of the important foundational questions about space-time arise in the context of tensor fields in Special Relativity and General Relativity, so one need not think about spinor fields to answer them.  It is also widely believed that such differences as spinor fields make are those described by Weyl in 1929, with some elaboration in terms of global methods.  To the contrary,  one of the most important developments for space-time theory since the rise of General Relativity appeared in a 1960s Russian journal due to the work of particle physicists wielding perturbative expansions.  That work can be unified with the classical differential geometry literature on nonlinear geometric objects.  The result is a formalism with fewer fields, fewer symmetries to gauge the extra fields away, and well defined Lie and covariant derivatives.  Fermions are more relevant to foundational questions about time and space-time theory, more like bosons in some respects, and less like bosons in others, than has been widely believed.


%
%
%



\end{document}